\documentclass[12pt,preprint]{aastex}
\usepackage{natbib}

\shorttitle{Age of Cluster Cooling Flows}
\shortauthors{Soker \& David }
\begin{document}

\title{Observed Non-Steady State Cooling and the 
Moderate Cluster Cooling Flow Model}

\author{Noam Soker\altaffilmark{1} and
Laurence P. David\altaffilmark{2} }
\altaffiltext{1}{ Department of Physics, Oranim, Tivon 36006, Israel;
soker@physics.technion.ac.il.}
\altaffiltext{2}{ Center for Astrophysics, Cambridge, MA 02138, USA;
david@cfa.harvard.edu}

\begin{abstract}

We examine recent developments in the cluster
cooling flow scenario following recent observations by
{\it Chandra} and XMM-Newton.  
We show that the distribution of gas emissivity verses temperature
determined by XMM-Newton gratings observations demonstrates that the
central gas,
{{{ i.e., where the cooling time is less than the age of the
cluster,}}} 
in cooling flow clusters cannot be in {{{ simple}}} 
steady-state, i.e.,
$\dot M$ is not a constant at all temperatures. 
Based on the measured gas emissivity, the gas can only be in 
steady-state if there exists a steady heating mechanism that
scales as $H(T) \propto T^{\alpha}$ where $\alpha=1-2$. That is,
a heating mechanism that preferentially targets the hottest and
highest entropy gas, which seems very unlikely.
Combining this result with the lack of spectroscopic evidence for
gas below one-third of the ambient cluster temperature is strong
evidence that the gas is heated intermittently.
While the old steady-state isobaric cooling flow model is incompatible with 
recent observations, a "moderate cooling flow model", in which the gas undergoes
intermittent heating that effectively reduces the 
age of a cooling flow is consistent with observations.
Most of the gas within cooling flows resides in
the hottest gas, which is prevented from cooling continuously and attaining 
a steady-state configuration.
This results in a mass cooling rate that decreases with decreasing
temperature, with a much lower mass cooling rate at the lowest
temperatures. Such a temperature dependent $\dot M$ is required
by the XMM-Newton RGS data and will produce an increasing amount
of intermediate temperature gas which will then be reheated during 
the next heating cycle.
We show the compatibility of this model for the cooling flow cluster A2052.
{{{The present paper strengthens the moderate cooling flow model,
which can accommodate the unique activities observed in
cooling flow clusters. }}} 

\end{abstract}

\keywords{
galaxies: clusters: general ---
cooling flows ---
intergalactic medium ---
X-rays: galaxies: clusters
}

\section{Introduction}\label{sec:intro}

Several recent papers have shown that the predictions of the
steady-state cluster cooling flow (CF) model are inconsistent
with {\it Chandra} and XMM-Newton X-ray observations
(see review by Fabian 2003).
Chandra observations show that the gas in the central regions
of relaxed clusters with central dominant galaxies can be described as
homogeneous single temperature gas with a positive temperature gradient.
Ettori (2002) showed that the ASCA evidence for multiphase 
CFs was due to the poor spatial resolution of ASCA which could not
distinguish between multiphase gas and single phase gas with
a temperature gradient.
Chandra observations show that only within the central few tens of kpc
in clusters does the spectroscopy require additional components above
a single temperature.
Of course, this could simply be due to statistical and spatial resolution
limitations in the Chandra data.
The strongest evidence against the steady-state isobaric CF 
model is the lack of observered line emission in XMM-Newton RGS
spectra from gas cooler than $\sim 1/3$ of ambient cluster
temperatures (e.g., Kaastra et al. 2001; Peterson et al.\ 2001;
Fabian 2003).  In the so called ``standard steady-state CF model'',
the age of a CF is assumed to be similar to the age of the
cluster; namely the gas has been cooling for a long time.
In the so called ``moderate CF model'' (Soker et al. 2001) heating
is intermittent and the hot gas is not in steady-state, in the sense
that its cooling time is longer than the time elapsed since the gas
was last heated.
Many reheating scenarios of the central gas in CFs 
have been proposed in the past with the aim of suppressing 
CFs altogether, or significantly reducing the average mass cooling rate
(e.g., Binney \& Tabor 1995; Tucker \& David 1997;
Ciotti \& Ostriker 2001; David et al.\ 2001; Quilis, Bower, \& Balogh 2001;
Br\"uggen, M. \& Kaiser 2001; Ruszkowski \& Begelman 2002; Nulsen et al. 2002). 
{{{These scenarios can be divided into models with steady
heating by conduction or AGN, or non-steady, self-regulating models heated 
by nuclear outbursts.  The moderate CF model is in the class 
of non-steady models where the gas cooling rate is peridocially
heated by nuclear outbursts. We present below the evidence for non-steady
CFs.}}}
The frequent occurrence of X-ray cavities coincident with radio lobes
around the central dominant galaxy in CFs
(e.g., Abell 496, Dupke \& White 2002; Perseus [Abell 426], Fabian et al.
2002; Hydra A, McNamara et al. 2000) demonstrates that AGNs have a
significant impact on the X-ray morphology of the hot gas.
However, there are still significant uncertainties in the detailed physics
of how the relativistic and thermal plasmas interact and how much heat is 
deposited into the hot gas.

In the moderate CF model where the intermittent heating is generated by 
AGN activity in the central dominant galaxy, it is expected that the 
amount of cooling gas increases sharply with increasing temperature.
Only at very low temperatures is there a steady-state situation
(i.e., where the cooling time is short compared to the time between
nuclear outbursts).
In Soker et al. (2001) we presented the arguments for a moderate
CF in which the actual mass cooling rate is significantly below that 
derived under the assumption of the standard model, but a non-steady CF still exists.
Soker et al. (2001) estimate that in the moderate cluster CF model
the required kinetic energy of the AGN is
$\sim 10^{47}~{\rm erg}~{\rm s}^{-1}$, and its strong activity
should last $\sim 10^7$ yr and occur every $\sim 10^9$ yr.
Only $\sim 1 \%$ of all CF clusters should be found during that stage.
In Cygnus A there is a strong radio source which heats the
intra cluster medium (Smith et al. 2002).
Wilson, Young, \& Smith (2003) estimate the mechanical power of
the jets in Cygnus A to be
$L_{\rm jet} \simeq 6 \times 10^{46} ~{\rm erg}~{\rm s}^{-1}$.
This is $\sim 100$ times larger than the radio emission,
and it is in the range required by the moderate CF model.

In recent analyses of XMM-Newton RGS data by Kahn et al. (2003) and 
Peterson et al. (2003; hereafter P2003) they find that there
is less gas than that predicted by the steady-state isobaric CF
model at all temperatures below the ambient gas temperature. Also, the
discrepancy increases with decreasing temperature.  There is not
just a deficit in gas below $\sim 1/3$ of the ambient cluster temperature,
there is a deficit of gas at all temperatures below the ambient temperature.
They conclude that their results are difficult to reconcile
with the newly proposed alternatives to the standard CF model,
including those that completely suppress radiative cooling with 
some form of steady-state heating, and that new physics may be required.
Our goal in this paper is to show that the XMM-Newton results 
are consistent with the expectations of the moderate CF model presented
in Soker et al.(2001).

We convert the differential luminosity as a function of temperature as
derived from the RGS data on cluster CFs into the distribution
of gas mass verses temperature in $\S 2$.
We also derive $\dot M$ as a function of temperature in this
section and show that the gas in CFs cannot be in steady-sate.
In section 3 we apply the moderate CF model to a recent Chandra observation
of A2052, and in $\S 4$ we summarize our main results.

\section{Distribution of Gas Mass with Temperature}
\label{distribution}

We show in this and the next sections that the distribution of gas
mass with temperature within cluster CFs found by P2003 can
be incorporated into a model with intermittent heating, such that the
effective age of the cooling gas is only $\sim 1-3 \times 10^9$ yr.

Based on RGS spectra of 14 CF clusters,
P2003 find that the variation in the differential luminosity with gas temperature 
can be characterized by the following expression: 
\begin{equation}
\frac {dL}{dT} = \frac {5}{2}
\frac{\dot M_{\rm ocf} k}{\mu m_p}
(\alpha +1) 
\left( \frac {T} {T_0} \right) ^{\alpha},
\label{eq:L2003}
\end{equation}  
where $k$,$\mu$, and $m_p$ have their usual meaning,
$\dot M_{\rm ocf}$ is the inferred mass cooling rate based on 
the assumptions inherent in the steady-state isobaric CF model, $T$ is the 
gas temperature, and $T_0$ is the
maximum, or ambient cluster temperature.
In general, the luminosity of gas cooling isobarically within a temperature 
interval $d T$ is:
\begin{equation}
dL = \frac {5}{2}
\frac{k}{\mu m_p} \dot M(T) d T.
\label{eq:dL2}
\end{equation}
In steady-state, the rate gas cools into a given temperature interval must equal 
the rate gas cools out of the same temperature interval.  In other words,
$\dot M (T)$ must be a constant and equal to $\dot M_{ocf}$.  Comparing equations
(1) and (2) shows that the gas can only be in steady-state if $\alpha=0$.
Fitting the RGS data on their sample of 14 CF clusters, P2003 find 
that $\alpha \sim 1-2$. 
{{{Within the old CF radius$-$where cooling time equals the
cluster age$-$most of the radiation comes in the X-ray band.
Only in the very inner region of $r \lesssim 10-30$~kpc
a significant fraction of the energy lost by the cooling gas
may be emitted in the optical and UV band
(e.g., Fabian et al.\ 2002; Soker, Blanton \& Sarazin 2003). }}} 

The mass cooling rate as a function of temperature can be written as:
\begin{equation}
\dot M(T) = \frac  {dM}{dT} \frac {dT}{dt}=
\frac  {dM}{dT} \frac {T}{\tau_{\rm cool}},
\label{eq:dotm2}
\end{equation}
where we define the cooling time to be:
\begin{equation}
\tau_{\rm cool} (T) \equiv \frac  {T}{dT/dt}.
\label{eq:dotm4}
\end{equation}
>From equations~(\ref{eq:dL2}) and (\ref{eq:dotm2}) we obtain
\begin{equation}
\frac{dL}{dT}  = \frac {5}{2} \frac{k}{\mu m_p} 
\frac  {dM}{dT} \frac {T}{\tau_{\rm cool}}. 
\label{eq:dL4}
\end{equation}
Combining equations~(\ref{eq:L2003}) and (\ref{eq:dL4}) gives:
\begin{equation}
\frac {dM}{dT} =
\dot M_{\rm ocf} \tau_{\rm cool} (\alpha +1)
\left( \frac {T} {T_0} \right) ^{\alpha-1} \frac{1}{T_0} .
\label{eq:dotm6}
\end{equation}
The cooling time varies as:
\begin{equation}
\tau_{\rm cool} =
\tau_0  \frac {\Lambda_0}{\Lambda} \frac {T}{T_0} \frac {n_0}{n} =
\tau_0  \frac {\Lambda_0}{\Lambda} \frac {P_0}{P}
\left( \frac {T}{T_0} \right)^2 ,
\label{eq:tcool}
\end{equation}
where $\tau_0$ is the cooling time of gas at $T_0$,
$n$ is the total number density, $P$ is the gas pressure,
and $\Lambda$ is the radiative cooling function,
such that $\Lambda n^2$ gives the energy radiated per unit volume per
unit time.  The last two equations can be combined to give:
\begin{equation}
\frac {dM}{dT} =
\dot M_{\rm ocf} \tau_0 (\alpha +1)
\left( \frac {T} {T_0} \right) ^{\alpha+1} \frac{1}{T_0} 
\frac {\Lambda_0}{\Lambda} \frac {P_0}{P} .
\label{eq:dotm8}
\end{equation}
The cooling function can be characterized as:
\begin{equation}
\frac{\Lambda}{\Lambda_0} = \left( \frac {T} {T_0} \right)^{\eta}
\end{equation}
where $\eta \simeq 1/2$ for $T \gtrsim 2 \times 10^7$ and
$\eta \simeq -1/2$ at lower temperatures.
For consistency with the expressions
derived above, assuming isobaric cooling and integrating 
equation~(\ref{eq:dotm8}) over temperature gives:
\begin{equation}
M(<T)=
\dot M_{\rm ocf} \tau_0
\frac {\alpha+1}{\alpha-\eta+2}
\left( \frac {T} {T_0} \right) ^{\alpha-\eta+2} 
\label{eq:m2}
\end{equation}
Setting $\alpha \simeq 1.5$ in the last equation, the average value found
by P2003, and assuming Bremsstrahlung cooling ($\eta =0.5$), gives:
\begin{equation}
M(<T) \simeq
\dot M_{\rm ocf} \tau_0
\left( \frac {T} {T_0} \right) ^3 . 
\label{eq:m4}
\end{equation}

If we add a heating mechanism to equation (2), then a steady-state
condition can be established only if $H(T) \propto T^{\alpha}$.
It is difficult to conceive of a heating mechanism that preferentially
heats the hottest and highest entropy gas in a CF.
The RGS data show that $\dot M(T)$  increases with increasing temperature.
If this were true over the lifetime of a cluster, there would
be a large reservoir of gas at intermediate temperatures,
which is also inconsistent with the RGS data.  This gas must be 
periodically removed from these intermediate temperatures either by
cooling sporadically to very low temperatures, which simply
returns us to the classic CF problem, i.e., the
lack of a significant reservoir of cool gas, or intermittent
heating back to roughly the ambient temperature. 
We therefore examine the non-steady moderate CF model.

\section{Moderate Cooling Flow Model}

As an example, we consider the CF cluster A2052, which was included 
in the P2003 XMM-Newton sample, and whose X-ray structure as observed
by Chandra was discussed in detail by Blanton et al. (2001, 2003).
P2003 assume a cooling radius of $r_0= 51$
arcsec, and find $\alpha \simeq 3$ for this cluster. Hence,
from equation~(\ref{eq:m2}), $M(<T) \propto T^{4.5}$, and
most of the mass is in the hottest gas.

Indeed, from fig. 4 of Blanton et al. (2003) we find
that the mass within 30 arcsec consists of $\sim 30 \%$ of
the total gas mass within 50 arcsec.
The temperature of the gas at 30 arcsec is 2.5 keV
(Blanton et al. 2003).
Using this along with an ambient temperature of $kT_0=3.3$ keV
(Blanton et al. 2003), implies that $(2.5/3.3)^{4.5}=29 \%$ of the
total gas mass up to $T_0$ resides at temperatures lower 
than $2.5$ keV.
Considering the uncertainties, these two numbers are in excellent agreement.
A small fraction of the gas still resides at lower temperatures,
presumably because the shock that heated the gas, say 
$\sim 10^9$ yrs ago, could not increase the cooling time of the lowest
entropy gas above the time between outbursts (Soker et al. 2001).

P2003 did not compare their XMM-Newton results directly
with Chandra data.  We find the cooling time at 30 arcsec, with $kT=2.5$ keV
and $n_e=0.02$ cm$^{-3}$ (Blanton et al. 2001), to be
$\tau_{\rm cool} (30 {\rm kpc}) = 1.5 \times 10^9$ yr.
Interior to this radius the intracluster medium is disturbed by two large radio bubbles
(Blanton et al. 2001).
In the moderate CF model, the time interval between intermittent energy
deposition is $1-4 \times 10^9$ yr, hence the gas is prohibited from
settling into a steady-state configuration at $r>30$ arcsec.
Although the gas continues to cool and its cooling time gets shorter, it is
unable to reach a steady-state before the next nuclear outburst.

Only within $ r\sim 30$ arcsec may the gas have reached a steady-state,
however the Chandra data show that it was recently disrupted by the 
two radio bubbles.  We therefore expect that the actual 
cooling rate is much smaller than that in the old (standard) CF model. 
Based on a spectral analysis of the Chandra data,
Blanton et al. (2003) find the mass cooling rate to be
$26 < \dot M <42 M_\odot$~yr$^{-1}$, which is $\sim 1/3$ of
the old value within $r \sim 140$ arcsec (Perres et al. 1998).
Taking the actual cooling radius to be the radius within which 
steady-state has been established based on the shorter age in the 
moderate CF model, the cooling rate will be even lower than the value found by
Blanton et al (2003), i.e., we argue for a mass cooling rate of
$\dot M \lesssim 10 M_\odot$~yr$^{-1}$.

{{{Although we only study one cluster as an example, we note:
(1) In the moderate CF model suggested by
Soker et al.\ (2001), the heating doesn't inhibit the
CF in the very inner region $r \lesssim 10$~kpc.
In this region, the gas continues to cool to temperatures
of $T\sim 10^4$~K, even after a heating event.
Only in the outer regions does the heating event prevent the gas
from cooling to low temperatures.
Therefore, we don't expect the gas to be isothermal.
The heating event can't heat the gas to extremely high temperature
either, because this requires AGN energy output much
larger than typical observed values.
(2) The moderate CF model thus predicts that $\alpha$ in eq. (1),
or $\alpha-\eta+2$ in equation (10) will not take
extreme values.  We can't predict the exact range of values of $\alpha$ in the
present study; this requires numerical simulations with variable
conditions in the intracluster medium before each heating event, the
energy supplied by the event, the time elapsed between events,
and other processes, e.g., mergers.
(3) Some CF clusters have central cooling times shorter than
those in A2052.  This does not pose a problem for the moderate CF model
since the central regions of clusters still harbor a CF.
(4) In light of the uncertainties, e.g., the temperature profile in the 
inner regions, the presence of X-ray cavities, and in the model parameters
mentioned in point (2) above, it does not warrant a more extensive
comparison with other clusters at this point.
Future work, in particular numerical simulations of heating
events, will include a greater comparison between the moderate
CF theory and cluster observations.}}} 

\section{Summary}
\label{sec:conclusion}

We show that the differential luminosity of the gas
in cluster cooling flows as a function of 
temperature, as derived from RGS XMM-Newton observations,
is inconsistent with steady-state cooling flow scenarios, 
but is consistent with non-steady heating or moderate cooling flow models.
The findings of P2003 imply that most of the gas in cooling 
flows resides in the highest temperature phase
(eqs. 10 and 11 above). Within the context of the 
moderate cooling flow scenario (Soker et al. 2001),
at these temperatures and densities the CF cannot
reach a steady-state, and if the intermittent heating
continues, it cannot attain such a state.
Therefore, the rate of gas cooling at high temperatures is
much higher than at lower temperatures.

We demonstrate the applicability of this model to the
cooling flow cluster A2052 (Sec. 3). We
argue for a mass cooling rate of $\lesssim 10 M_\odot$~yr$^{-1}$
in this cluster, which is compatible with the upper limit found 
by P2003 for lowest temperature gas.
This is much lower than the cooling rate of hotter gas,
$\gtrsim 100 M_\odot$~yr$^{-1}$, found by (P2003) or
deduced in the old CF model by Peres et al. (1998),
and somewhat lower than the cooling rate derived recently
by Blanton et al. (2003) of $\sim 26-42 M_\odot$~yr$^{-1}$.
Overall, intermittent heating in the moderate CF model
(Soker 2001; Fabian 2003), or more frequent heating
( Blanton et al. 2003 for A2052), may account for the P2003 findings
without invoking new processes.
The intermittent heating model, with time intervals between
major heating events of $\sim 1-3 \times 10^9$ yr has the
advantage that no fine tuning is required to balance heating
and cooling, since the gas is heated to relatively high temperatures,
and then starts cooling. Most of the heated gas does
not cool to low temperatures before the next major heating event.
{{{Different values of the physical parameters, e.g., energy input and
time intervals between heating events, as well as a cluster's
properties, will give different values of $\alpha$ in
equations (1) and (10). Indeed, a large range of
values is observed, $\alpha \sim 1-3$, hinting on a wide
range in the physical parameters mentioned above, such
that no fine tuning is observed or required.}}} 
As shown in Soker et al. (2001), it is difficult to prevent the
lowest entropy gas from complete cooling (i.e., increasing the cooling time 
above the time between heating events).
Hence, a low $\dot M$ cooling flow can be sustained in the very central
region of clusters.

{{{ This paper strengthens the moderate cooling flow model
(Soker et al.\ 2001), by supporting the claim that the unique
activities observed in cooling flow clusters (e.g., McNamara 2002
and references therein), can be accommodated within its framework.
The problems which are need to be solved, e.g., the exact nature
of the heating events, are less severe than the crisis encountered
in the old cooling flow model.}}} 

\acknowledgements
{{{ We thank an anonymous referee for useful comments. }}} 
This research  was supported in part by grants from the
US-Israel Binational Science Foundation and GO2-3171.


\begin{references}

\reference{} Binney, J., \& Tabor, G. 1995, MNRAS, 276, 663

\reference{} Blanton, E. L., Sarazin, C. L., \& McNamara, B. R.
2003, ApJ, in press (astro-ph/0211027) 

\reference{} Blanton, E. L., Sarazin, C. L., McNamara, B. R., \&
Wise, M. W. 2001, ApJ, 558, L15 %  (BSMW) 

\reference{} Br\"uggen, M., \& Kaiser, C. R. 2001, Nature, 418, 301

\reference{} Ciotti, L., \& Ostriker J. P. 2001, ApJ, 551, 131

\reference{} David, L. P., Nulsen, P. E. J., McNamara, B. R.,
 Forman, W., Jones, C., Ponman, T., Robertson, B., \& Wise, M.
 2001, ApJ, 557, 546

\reference{} Dupke, R., \& White, R. E., III 2002, in ASP Conf. Ser. 262, 
High-Energy Universe at Sharp Focus: Chandra Science, ed. 
E. M. Schlegel \& S. Vrtilek (San Francisco: ASP), 51

\reference{} Ettori, S. 2002, MNRAS, 330, 971

\reference{} Fabian, A. C. 2003,  in Galaxy Evolution: Theory and
Observations, eds. V. Avila-Reese, C. Firmani, C. Frenk, \& C. Allen,
RevMexAA SC, in press (astro-ph/ 0210150)

\reference{} Fabian, A. C., Celotti, A., Blundell, K. M.,
Kassim, N. E., \& Perley, R. A. 2002, MNRAS, 331, 369 

\reference{} Kaastra, J. S., Ferrigno, C., Tamura, T., Paerels, F. B. S.,
     Peterson, J. R., \& Mittaz, J. P. D 2001, A\&A, 365, L99 

\reference{} Kahn, S. M., Peterson, J. R., Paerels, F. B. S., Xu, H.,
    Kaastra, J. S., Ferrigno, C., Tamura, T., Bleeker, J. A. M., 
    \& Jernigan, J. G. 2003,   (astro-ph/0210665)

\reference{} McNamara, B. R. 2002, The High-Energy Universe at 
Sharp Focus: Chandra Science, Symposium at the ASP meeting,
ASP Conference Proceedings, Vol. 262, eds. Eric M. Schlegel and
Saeqa Dil Vrtilek, 351

\reference{} McNamara, B. R., et al. 2000, ApJ, 534, L135

\reference{} Nulsen, P.E.J., David, L., McNamara, B., Jones, C., Forman, W. ,
	   Wise, M. 2002, ApJ, 568, 163.

\reference{} Peres, C. B., Fabian, A. C., Edge, A. C., Allen, S. W.,
     Johnstone, R. M., White, D. A. 1998, MNRAS, 298, 416

\reference{} Peterson, J. R.,  et al.\ 2001, A\&A, 365, L104

\reference{} Peterson, J. R., Kahn, S. M., Paerels, F. B. S.,
    Kaastra, J. S., Tamura, T., Bleeker, J. A. M., Ferrigno, C.,
    \& Jernigan, J. G. 2003, ApJ, submitted (astro-ph/0210662) (P2003)

\reference{} Quilis, V., Bower, R. G., \& Balogh, M. L.
    2001, MNRAS, 328, 1091

\reference{} Ruszkowski, M., \& Begelman, M. C. 2002, ApJ, 581, 223
   (astro-ph/0207471) 

\reference{} Smith, D. A., Wilson, A. S., Arnaud, K. A., Terashima, Y.,
   \& Young, A. J. 2002, ApJ, 565, 195 

\reference{} Soker, N., White, R. E. III, David, L. P., \& McNamara, B. R.
    2001, ApJ, 549, 832 
 
\reference{} Tucker, W., \& David, L. P. 1997, ApJ, 484, 602

\reference{} Wilson, A. S., Young, A. J., \& Smith, D. A. 2003,
    in Active Galactic Nuclei: from Central Engine to Hot Galaxies,
    ASP Conf. Ser., eds. S. Colin, F. Combes, and I. Shlosman
     in press (astro-ph/0211541) 

\end{references}
\end{document}